\documentclass[journal]{IEEEtran}
\usepackage{color}
\usepackage{graphicx}               
\usepackage[cmex10]{amsmath}
\usepackage{url}

\ifCLASSINFOpdf
\else
\fi

\newcommand{\rand}{$\delta$}
\newcommand{\clust}{$\alpha$}
\newcommand{\abin}{$\alpha$-BiN}
\newcommand{\nagent}{$t$}
\newcommand{\nplace}{$N$}
\newcommand{\navgcon}{$\mu$}
\newcommand{\buftime}{$p$}
\newcommand{\vthr}{$v$}
\newcommand{\giantdtn}{$G_d$}
\newcommand{\giantbnw}{$G_b$}

\newcommand{\G}{$\mathcal G$}
\newcommand{\Pl}{$P$}
\newcommand{\GP}{$\mathcal G_P$}
\newcommand{\GPstar}{$\mathcal G_{P}^{*}$}
\newcommand{\A}{$A$}
\newcommand{\p}[1]{$p_{#1}$}

\newcommand{\att}[1]{$\theta_{#1}$~}

\newcommand{\SI}{$S_1$}
\newcommand{\SII}{$S_2$}
\newcommand{\EI}{$E_1$}
\newcommand{\EII}{$E_2$}
\newcommand{\EIII}{$E_3$}
\newcommand{\RI}{$r_1$}
\newcommand{\RII}{$r_2$}
\newcommand{\RIII}{$r_3$}

\begin{document}
%
\title{Coverage analysis of information dissemination in throwbox-augmented DTN}
%
%
%

\author{Sudipta~Saha, 
        Animesh~Mukherjee~
        and~Niloy~Ganguly 
\thanks{Sudipta Saha, Animesh Mukherjee and Niloy Ganguly are with the department
of Computer Science \& Engineering, Indian Institute of
Technology, Kharagpur - 721302, India.
}}

\maketitle

\begin{abstract}

This paper uses a bi-partite network model to calculate the coverage
achieved by a delay-tolerant information dissemination algorithm
in a specialized setting. The specialized Delay Tolerant Network
(DTN) system comprises static message buffers or throwboxes kept
in popular places besides the mobile agents hopping from one place
to another. We identify that an information dissemination
technique that exploits the throwbox infrastructure can cover only
a fixed number of popular places irrespective of the time spent.
We notice that such DTN system has a natural bipartite network
correspondence where two sets are the popular places and people
visiting those places. This helps leveraging the theories of
evolving bipartite networks (BNW) to provide an appropriate
explanation of the observed temporal invariance of information
coverage over the DTN. In this work, we first show that
information coverage can be estimated by the size of the largest
component in the \textit{thresholded one-mode projection} of BNW.
Next, we also show that exploiting a special property of BNW, the
size of the largest component size can be calculated from the
degree distribution. Combining these two, we derive a closed form
simple equation to accurately predict the amount of information
coverage in DTN which is almost impossible to achieve using the
traditional techniques such as epidemic or Markov modeling. The
equation shows that full coverage quickly becomes extremely
difficult to achieve if the number of places increases while
variation in agent's activity helps in covering more places.


\end{abstract}


%
\IEEEpeerreviewmaketitle

\section {Inroduction}
Delay Tolerant Networks are decentralized (in their primitive
version peer-to-peer) communication systems consisting of multiple
handheld devices carried by mobile agents (humans). One of the
fundamental components of any DTN system comprises efficient
message dissemination.
With the advancement of DTN protocols, stationary message storing
devices like `relay points' or `throwboxes'
\cite{DTN-info-disse-Throwbox} are now being introduced to
artificially and indirectly increase the contact opportunities
among the mobile devices. Generally, these devices are installed
in common places where people usually move, such as markets, pubs,
public gatherings, relief centers during disaster etc. Mobile
devices drop and collect messages from these throwboxes. This
gives rise to a special kind of dissemination process where, a
piece of message passes from throwbox (place) to throwbox (place)
and indirectly get transferred to people (mobile devices). Since
throwbox forms the basis of information dissemination, it is
important to observe the number of distinct throwboxes ultimately
receiving a distinct piece of message as that would indirectly
indicate the extent to which the message reaches the  mobile
devices \cite{SimilarWork-recent-journal}. The number of
throwboxes which has received and hosted the piece of information
is termed as {\em place coverage} throughout the paper.
{\em In this work, we set the computation
of this place coverage under various parameter settings, as the primary objective}. Note that this
problem is slightly different from the traditional spreading
problem where one is mostly interested in the number of
`\textit{infected/occupied}' individuals at a particular time; in
contrast, here we are interested in the number of place that have
the message in the asymptotic limits.

In order to understand the dynamical nature of {\em
place coverage}, we perform extensive simulation, considering both
artificial and real settings (an usual policy adopted for such
studies in the literature \cite{SimilarWork-recent-journal}). We
observe 
that non-intuitively the {\em place coverage}
stabilizes after a certain period of time. In order to explain
this phenomenon as well as to systematically understand the
influence of each parameter, a mathematical model is necessary. In
this context, we identify that the message dissemination process
in a message buffer augmented DTN has a natural and one-to-one
correspondence with a growing bipartite network
\cite{BIP-Saptarshi,2004_informationprocessingletters_jean_bnw}
where one partition contains a fixed number of places and the
other partition contains the agents whose number continuously
grows over time. We precisely attempt to show that, the existing
rich theoretical backbone of evolving bipartite networks
\cite{BNW_others_pre,BIP-choudhury-2010-81,RandomIntersectionGraphs,HiddenBNW,BIP-Mukherjee-1,BIP-Peruani-1}
can be exploited to analyze the otherwise intractable
characteristics of message dissemination process in DTN. The
primary contributions of this work are as follows -

(a) We show that the information coverage can be estimated using
the size of the largest component of the thresholded projection of
the growing bipartite network appropriately constructed from the
DTN.

(b) We also show that due to a special property of such bipartite
networks, this largest component size can be easily derived from
the degree distribution of the thresholded projection. Using this
technique, we derive a closed form simple equation which
faithfully captures {\em place coverage} in DTN.

(c) Finally, we draw several insights regarding the complex
stochastic dynamics of {\em place coverage} in the indirect
message transfer process in DTN. We find that the it ({\em place
coverage}) decreases sharply (at a rate proportional to the cube
of the number of places) with the increase in the number of places
to be visited. Furthermore, it increases linearly when the
mobility pattern becomes a bit more random. We also find that the
variation in agents' activity indirectly helps in enhancing this
{\em place coverage}.

The rest of the paper is organized as follows. In the next
section, we precisely describe the basic setup and the protocol
followed in disseminating information in DTN using the installed
buffers at the common places. Subsequently, in section
\ref{sec:BipartiteNetworkEvolution}, we describe the corresponding
BNW model of the DTN and the analytical estimate of the achieved
coverage in DTN using the size of the largest connected component
of the thresholded one-mode projection of the BNW. In section
\ref{sec:insights}, we show the influence of various parameters
such as the number of places in the system, the activeness of the
agents etc. through the analytical framework. In this section, we
also describe the effect of incorporating the presence of
randomness as well as the clustering exponent in the selection of
the next place to be visited by the mobile agents. In section
\ref{sec:RelatedWork} we present a brief review of the
state-of-the-art before drawing the conclusion. We have presented
some very preliminary results in a workshop
\cite{DTN-buffer-mobiopp}; the state-of-the art section clearly
notes the radical improvement made in this paper.

\section{Information dissemination in DTN}\label{sec:InformationDisseminationDTN}

In this section we first describe the scheme employed to
disseminate information in DTN exploiting the presence of the
message buffers in various common places. We perform detailed
experimentation to find out how the place coverage in {\em
throwbox augmented DTN} increases with time. Surprisingly we find
that the achieved place coverage stabilizes with time. We further
validate the results with some publicly available GPS trace based
data. We also show that the {\em place coverage} and the {\em
agent coverage} bear a strong positive correlation.

\subsection{Basic settings}

The social movement patterns of the humans play a very important
role in the performance of any information dissemination algorithm
implemented in DTN. Therefore, we adopt the \textit{Clustered
Mobility Model} (CMM) \cite{CMM}, which primarily captures this
social movement patterns (details in section
\ref{subsec:placeselection}). Following the principles of CMM
\cite{CMM}, we consider that in a certain geographic area there
are a fixed and limited number of \nplace{} common places and each
of them is equipped with message buffer(s). We also assume that
each agent carries a mobile device which can store message in
its local buffer and is also able to send/receive message
to/from the throwbox installed at the common places. Detailed
description of the actual message dissemination
process, is given in the next subsection.

\subsection{The dissemination process}\label{subsec:dissemination}
\noindent \textbf{Initiation}: The dissemination process is
initiated by a single agent who wants to spread a particular piece
of message to other agents. The initiator agent creates/contains
the message (with unique message identifier) and carries the same
to different places as he visits those places. Multiple messages
with distinct message identifiers, can be initiated by multiple or
same agent. However, we consider here only a single agent is
disseminating a single message. As the agent comes to a place, a
copy of the message is dropped in the buffer(s) installed at that
place. When the message is dropped to some place in this way, that
place also takes part in the dissemination process as described
below.\newline \textbf{Role of a place}: When an agent arrives at
a place where the initiator agent has already dropped the message,
it gets transferred to the agent's device, if the device is not
already carrying the message.\newline \textbf{Role of other
agents}: The role of the other agents, starts after they pick up
the message from some place. They behave similarly as the
initiator, i.e., drop copies of the collected message to different
places as they visit.\newline \textbf{Buffer refreshment}: It is
to be noted that, the message buffers have limited capacity.
Therefore, in order to remove stale messages from the system, as
well as to disseminate multiple messages simultaneously, some of
the existing messages in each of the buffers require to be deleted
periodically. To facilitate this, we assume a global {\bf refresh
probability} (denoted by \buftime) which is pre-decided and at each
time step (following the physical clock), each one of the distinct
messages in each of the message buffers is deleted using refresh
probability \buftime.

{\em Practical value of p}: It is to be noted that, if the applied
refresh probability is close to zero, then full coverage would be
ensured and hence is uninteresting for any further investigation.
Also, in all practical cases the value of \buftime, cannot be too
high - which would imply too little coverage. Therefore, we have
kept the range for $p$ from 0.01 to 0.2 in our study.

\textbf{No direct communication}: It is to be noted here that we
only consider the indirect/asynchronous communications among
agents via the installed buffers in the common places. In other
words, direct agent to agent message passing is not considered in
the model.

\begin{figure*}[htbp]
\begin{center}
\includegraphics[angle=0,width=\textwidth, height = 5cm]{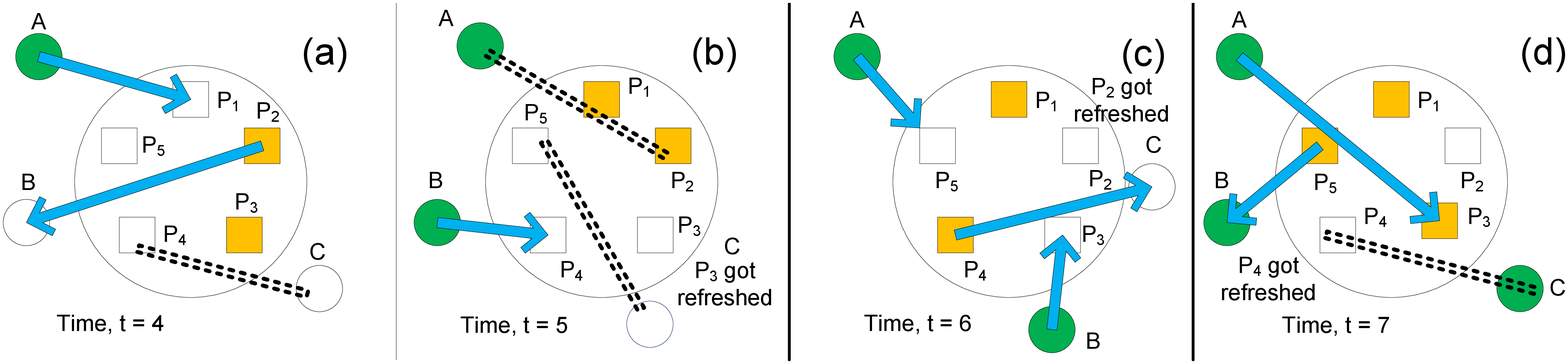}
\end{center}
\caption{A schematic diagram showing the information dissemination
process running in a DTN. Parts (a), (b), (c) and (d) show the
status of the system at the beginning of the $4^{th}$, $5^{th}$,
$6^{th}$ and $7^{th}$ time steps, in the process of dissemination.
There are 5 places: $P_1$ ... $P_5$ and 3 agents $A$,$B$,$C$ in
the system. Each of the agents makes one visit to one of the
places at every time step. We consider the dissemination of a
single message only and hence, we consider that there is a buffer
in each of the agents and the places which can hold a single
message. The empty and the filled up circles (squares) denote
agents (places) without or with the message in their buffers,
respectively, at the specific annotated time step. As an agent
makes a visit to a certain place, transfer of message happens only
in the following two cases - (i) place has the message and the
agent does not have it, or (ii) agent has the message but the
place does not have it. The arrows denote the direction of
transfer of the message. The dotted lines indicate that no message
transfer took place. At the end of each time step, all the buffers
are refreshed (message deleted) according to some probability
value. The changes due to an operation performed in a particular
time step is reflected in the next time step.}
\label{fig:DTNmodel}
\end{figure*}

A sample scenario for such kind of dissemination strategy has been
pictorially described in figure \ref{fig:DTNmodel}. In this
example we have assumed that there are three agents in the system
and at every time step each one of them makes a visit to some
place. The buffers are probabilistically refreshed at the end of
every time step.

\subsection{Metrics}

In our analysis, we use the two metrics as defined below to
measure the efficiency of the dissemination process.

\textit{Agent coverage:} The number of agents whose buffers
contain the message that is being disseminated.

\textit{Place coverage:} Given a message, the total number of
throwboxes\footnote{Throwbox and place are used synonymously.}
whose buffers receive and host the message (need not be
simultaneously)  due to the process of indirect dissemination
described above.

In the small example depicted in figure \ref{fig:DTNmodel}, the
agent coverage values at the $4^{th}$, $5^{th}$, $6^{th}$ and
$7^{th}$ time steps are 1, 2, 2 and 3 respectively whereas the
place coverage values at these time steps are 2, 2, 2 and 3
respectively. In this work, we mostly concentrate on the place
coverage. However, we give an analysis of both of these metrics in
section \ref{subsec:placevsagentcoverage}.

There are two major issues associated with the described
dissemination process. The first one is the agents' mobility and
the second one is the agents' activity. In the next two
subsections, we discuss the effect of both of these issues on the
achieved coverage.

\subsection{Agents' mobility} \label{subsec:placeselection}

In this work the agents are considered to be only humans. In
social context, preferential visit to different locations by
humans has been the central idea of many proposed mobility models
\cite{mobility_swim}. However, in order to keep it simple, in our
analysis we consider the generalized version of the preferential
selection process (complying with CMM) described as follows. At
time step $t$, the chance that a place $P_i$ gets visited next by
an agent, is modeled by the probability calculated through the
following expression:
\begin{equation}
\label{eq:pref_rand_probability} Prob(P_i) =
\frac{\left(d_i(t)+\delta\right)^\alpha}{\sum_{j=1}^{N}\left(d_j(t)+\delta\right)^\alpha}
\end{equation}

\noindent where $d_i(t)$
reflects the number of agents that have already visited the place
$P_i$ by time $t$ (i.e., the preference factor associated with a
place); the parameter \clust{} is the clustering exponent
\cite{CMM}. In order to separately study the effect of the
existing randomness (that moderates the preference factor) in the
mobility of the agents, we introduce a parameter \rand{} ($\ge
0$). We also assume that the stay times of the agents in the places
are high and hence we ignore the transition time taken by an agent
to jump from one place to the other
\cite{SimilarWork-recent-journal}.

However, in order to focus on the more fundamental properties of
the dissemination process, we mostly consider \rand{} = 0, and
\clust{} = 1 for which the system reduces to a pure preferential
one. Later, in section \ref{sec:insights} we discuss the effects
of the non-zero values of \rand{} and \clust.

\subsection{Agents' activity}

From a global perspective, it may seem that some visits by
different agents in the system take place simultaneously, i.e., at
the same time. However, a more granular observation of the events
reveal the fact that the probability that two distinct visits by
two distinct agents to two different places or even at a same
place happening exactly at the same time, is very low. In other
words, at micro level timescales it is possible to perfectly order
the events one after the other. Hence, without loss of generality,
we assume that the agent's visits happen sequentially and hence
can be serialized according to the time stamps of those visits.
This serializability of the sequence of agent visits allows us to
rearrange them in a desired fashion whereby at each time step it
is assumed that there is exactly one agent visit. We, however,
assume that always the first agent in the system brings the
message to be disseminated.

\begin{figure*}[htbp]
\begin{center}
\includegraphics[width = 0.8\textwidth, height = 8cm]{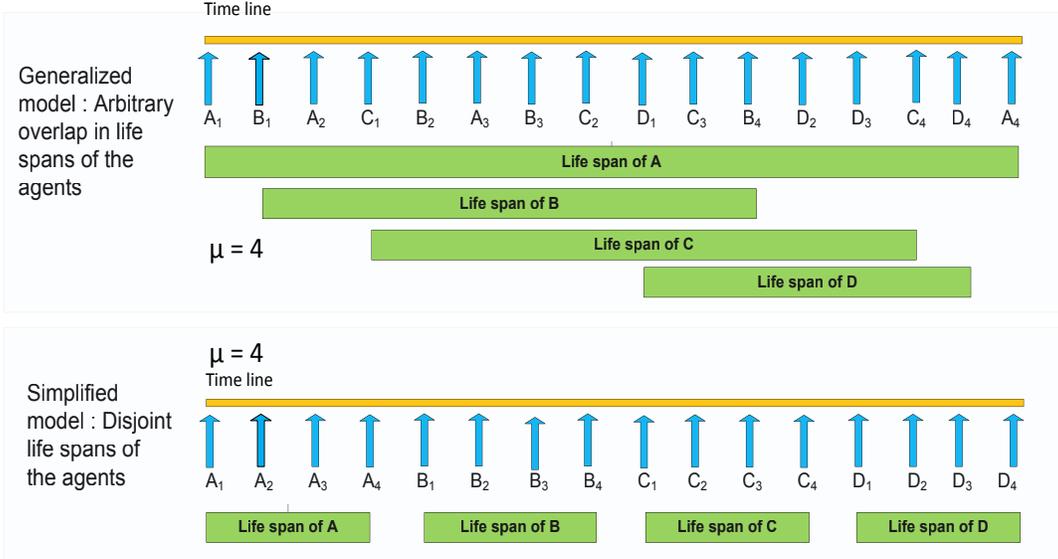}
\end{center}
\caption{Disjoint and overlapping life spans of the agents. Four
agents have been considered which are denoted by - A, B, C and D.
Each agent is assumed to visit 4 places in the system (i.e.,
$\mu$=4). The subscript $i$ after an agent $X$ denotes the
$i^{th}$ visit by the agent $X$. Part (a) shows the most general
case where the life spans of the agents are overlapping with each
other. Potentially any agent could have been active till the end. Part (b) shows the case where the life spans are all
disjoint with each other.
}
\label{fig:lifespan}
\end{figure*}

The duration of time that an agent actively participates in the
dissemination process by visiting various places, can be termed as
its \textit{life span} (in other words, it is the time gap between
the agent's first visit and his last visit). These life spans of
different agents may overlap with each other. The degree of this
overlap can be controlled by restricting the life spans of the
individual agents on the time line. In an extreme case of this
spectrum, we have a fully overlapping model of the agent's
activity where the life spans of the agents may arbitrarily
overlap with each other. In other words, if we consider the total
period of activity as $T$, then in the said extreme case, an agent
(say $n$) potentially stays active for the time span $E_n$ to $T$,
where $E_n$ is the entry time of the agent $n$. The other extreme
case is the model where the life spans of all the agents are fully
disjoint from each other that is the total period of the activity
of an agent is $E_n$ to $E_n + \mu -  1$ where $\mu$ is the number
of places visited by it. In between these two cases, we can
appropriately restrict the life spans of the agents to reflect the
cases where the life spans partially overlap with each other.
These two scenarios of agents' life span are pictorially explained
in figure \ref{fig:lifespan}.

\begin{figure*}[htb]
\begin{center}
\includegraphics[width=0.8\textwidth, height=7cm]{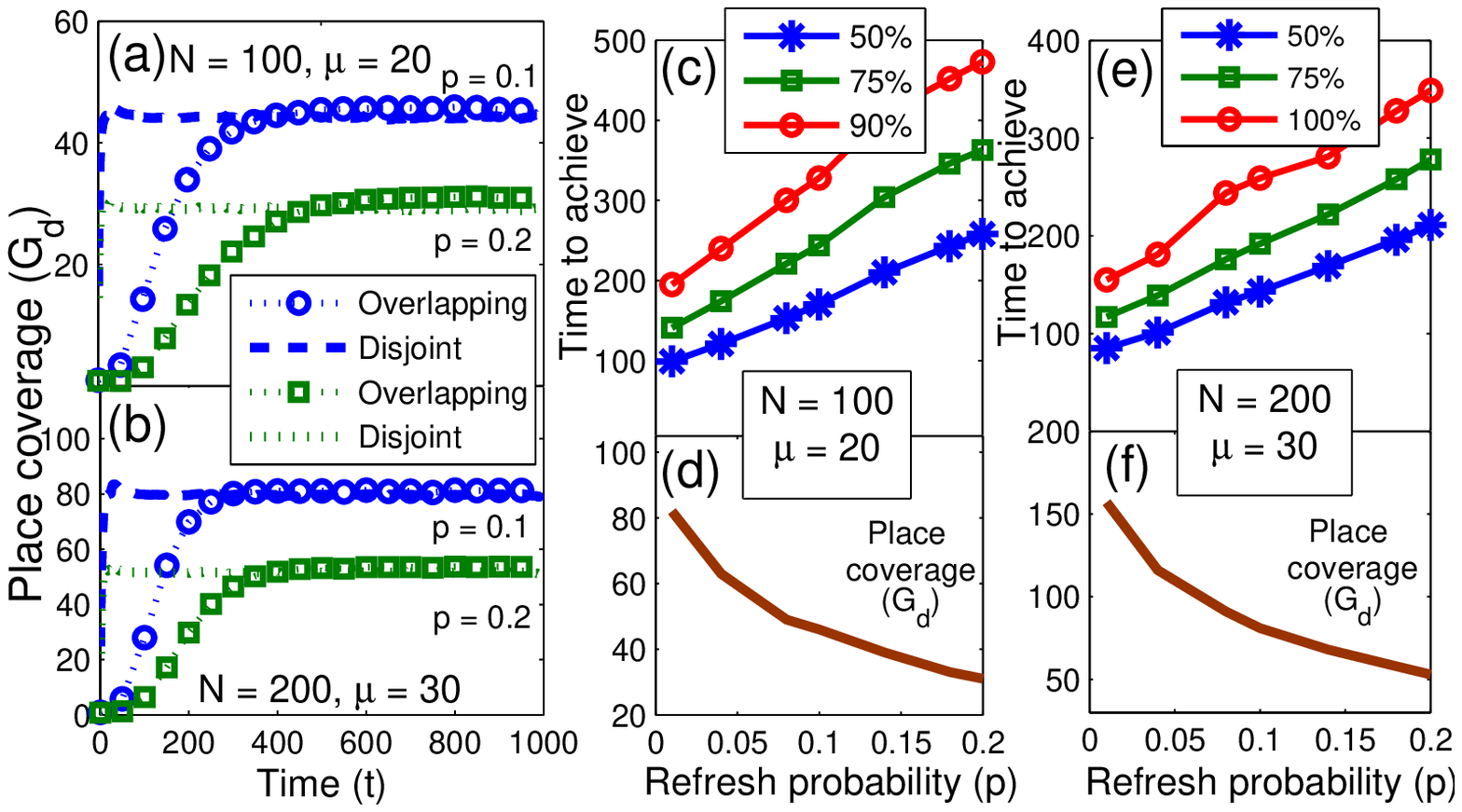}
\end{center}
\caption{Parts (a) and (b) show the time evolution of the achieved
coverage in the simulation of the dissemination process under
overlapping and disjoint life spans of the agents for two
different refresh probabilities and two different combinations of
\nplace{} and \navgcon. Parts (c) and (e) show the plot of the
relationship of the time to achieve a certain percentage of agent
coverage with the applied refresh probability \buftime{} for two
different combinations of \nplace{} and \navgcon{}. Parts (d) and
(f) show the plot of the achieved place coverage for the refresh
probabilities used in parts (c) and (e), respectively. In all these
simulations, number of agents has been assumed to be 1000 and the
results shown are the average over 1000 runs.}
\label{fig:sequential-eq-overlap}
\end{figure*}

An important aspect to observe is the difference in performance
arising due to varying degree of overlap. To check the difference,
we empirically compare the time evolution of the achieved coverage
under the two extreme cases of agents' activity - fully
overlapping and fully disjoint. To make a fair comparison, we
assume that at each time step, in both the cases, \navgcon{}
distinct visits take place. Figure \ref{fig:sequential-eq-overlap}
shows the results for various combinations of \nplace{} and
\navgcon. In all the cases, we find that for a given refresh
probability (\buftime), the coverage gets stabilized to a same
value after a certain number of visits (initial delay) have been
completed by the agents. The initial delay to reach the stability
is much higher in the overlapping case. However, the significant
observation obtained through experimentation is that the {\em
overlaps in the life spans of the agents do not have any specific
effect} on the final calculation of {\em place coverage}.


We find that the final stabilized place coverage achieved in the
dissemination process solely depends on the system parameters
\nplace, \navgcon{} and the applied value of \buftime. Hence, we
concentrate only on this stabilized value of the coverage and its
relationship with the other parameters. In order to get rid of the
additional complexities due to the overlaps in the life spans of
the agents, we adopt a simplified strategy described as follows.
We assume that the life spans of the agents are disjoint. We also
assume that each of the agents enters one by one, makes \navgcon{}
number of visits to different places and exits the system. This
parameter \navgcon{} actually models the average social
participation of the agents. Here we have taken it as a constant
but, analytically it can be shown \cite{BIP-Saptarshi} that the
results remain same if it is assumed to be the average of the
distribution of the number of visits an agent makes in its whole
life span. Moreover, exploiting this simplistic setup we also
safely assume that the dynamical time of the system is advanced
after each individual agent completes all its $\mu$ visits as well
as the buffers are refreshed (with probability $p$) at the end of
every time unit.


\subsection{Place coverage versus agent
coverage}\label{subsec:placevsagentcoverage}

It should be noted that, for the sake of simplicity, in this
analysis unlike the place buffers we do not consider any
refreshing / resetting of the agents' buffers. Therefore, with
time the agent coverage gradually increases and after a certain
time, all the agents that participate in the dissemination process
get the message. Hence, the more important metric is to check the
time needed to cover all the agents.
We have already observed that place coverage decreases with the
enhancement of refresh probability (figures
\ref{fig:sequential-eq-overlap} (a) and (b), also (d) and (f)).
Figures \ref{fig:sequential-eq-overlap} (c) and (e) show the
relationship between the time to achieve a few different
percentage of agent coverage with the applied refresh probability.
The effect of the refresh probabilities is clearly visible - for
higher values of the refresh probability, the time it takes to
achieve a certain percentage of agent coverage is also high. {\em
Hence higher place coverage directly implies faster agent
coverage. } Thus, considering the importance of this place
coverage, we set it to be the prime target of all the analysis
presented in this paper. In the rest of this paper, we use the
word `coverage' to implicitly refer to this `place coverage'
(`agent coverage' is explicitly mentioned). We denote this
quantity by \giantdtn.

\subsection{Real data analysis}

We use few publicly available GPS traces \cite{crawdad} to test
whether the empirically observed temporal stability in coverage,
also appears in real situation. In order to test this hypothesis
the GPS trace needed to be appropriately processed which is
elaborated next.

{\bf Data processing:} These GPS trace data contain the
information of the visits of a small number of agents at different
places at different times within a certain campus area. However,
the data of agent's movement were of different days making it very
sparse. Assuming regular pattern of activity where a trace in day
$x$ would also appear in day $y$, we considered the trace record
of a given agent for  different days to be the records of separate
agents. We subsequently merged all the trace records (of all the
agents) together producing a consolidated trace record of a single
day. We create equi-sized circles around different trace points
and term the circular zones as places. We ignore the consecutive
visits within same circle by the same agent. However, if an agent
comes back to within a certain circle after visiting at least one
different circle, we consider it to be a distinct visit. After
obtaining all the visit points, we sort these common places in a
descending order according to the number of visits to the places
by different agents. To extract out the most popular places, we
take only the top 20 places from this sorted list. This way we
prepare four different datasets - KAIST, NC State fair, New York
and NCSU (for details of the datasets, please see \cite{crawdad}).

{\bf Experiment:} We run the dissemination algorithm already
described on these processed datasets. We assume the existence of
the message buffers (throwboxes) in the common places. We do the
refreshing of the buffers and measure the coverage after the
completion of every $k$ number of distinct visits by the agents.
The value of $k$ is determined for each dataset depending on the
total temporal length (number of hours) of the dataset as well as
the frequency of the agent visits. For KAIST, NC State fair, New
York and NCSU the values of $k$ are 50, 10, 4 and 4, respectively.

\begin{figure}
\begin{center}
\includegraphics[width=0.5\textwidth, height=7cm]{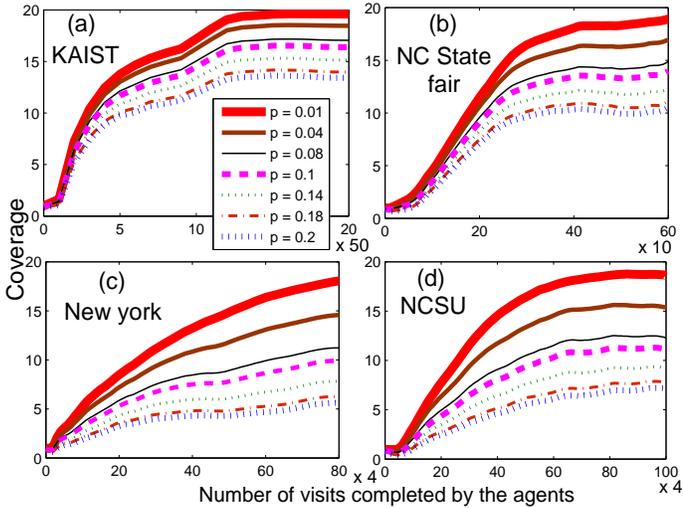}
\end{center}
\caption{Plot of the coverage achieved by the application of
message buffer in the information dissemination process
implemented on four real datasets (a) KAIST, (b) NC  State fair,
(c) New York and (d) NCSU. The shown results are the average over
1000 runs of simulation of the dissemination process.}
\label{fig:real}
\end{figure}

{\bf Results:} The results are presented in figure \ref{fig:real}.
It can be observed from all these results that, for almost all
different refresh probabilities the achieved coverage gets
stabilized after a certain initial number of visits by the agents
are over. To conclude, such stabilization is universal and is not an
 artifact of artificial systems. Hence,
understanding the basic reason for this stabilization is an
important step towards understanding the coverage dynamics.

\subsection{Modeling with bipartite networks (BNW)}
A natural and obvious interpretation of the underlying structure
of DTN is a bipartite network where one partition (set of popular
/ common places) remains fixed in size over time and the other
partition (set of agents and their visits) grows over time. Recent
studies on BNW, such as \cite{BIP-Saptarshi, BIP-JSac}, have
termed this kind of structure as \abin. These works reveal few
time invariant properties of such kind of BNWs. Being motivated by
these works, we hypothesize that \abin{} can be the ideal
candidate to explain the stabilization of the place coverage
(\giantdtn{}) achieved in the DTN settings. To investigate further
in this direction, we first formally model DTN as a BNW. We
precisely establish a one-to-one mapping of the quantity which has
to be analyzed in the DTN domain, as a quantity in the BNW domain.
We briefly describe these techniques in the next section.

\section{\abin{} and its relationship with DTN}
\label{sec:BipartiteNetworkEvolution}

\begin{figure*}
\begin{center}
\includegraphics[angle=0,width=\textwidth,height=5cm]{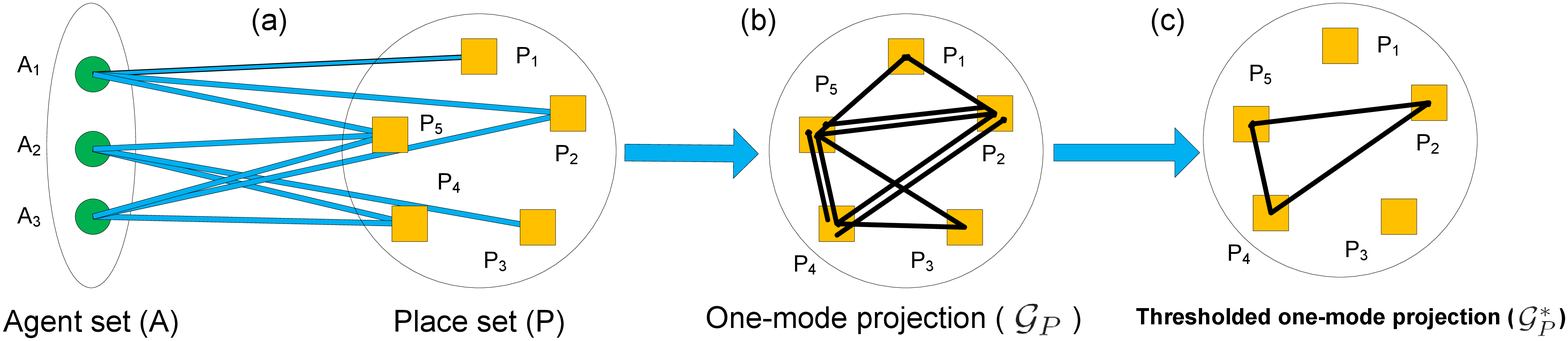}
\end{center}
\caption{A schematic diagram of a possible scenario of the BNW
corresponding to a DTN comprising five places, i.e., \nplace=5
($P_1$...$P_5$) and three agents, i.e., \nagent=3 ($A_1$ ...
$A_3$) with \navgcon=3. The diagram shows a possible status after
all the agents have joined the system. Part (a) shows the BNW, (b)
is the one-mode projection and (c) is the thresholded one-mode
projection for threshold value 2.} \label{fig:BIPModel}
\end{figure*}
In order to establish the relationship between \abin{} and DTN,
several basic terminologies need to be defined which is done next.
We then establish the correspondence.

\subsection{\abin~and its thresholded projection}
As mentioned, \abin{} is a bi-partite network where one partition
(the place partition) is fixed and finite while the number of
agents (the other partition) grows over time and is modeled by the
parameter \nagent. At any time instant $t$, an agent joins \abin{}
with \navgcon{} connections where the connections are chosen
preferentially (see figure \ref{fig:BIPModel}). This can be
thought of as the DTN situation where the life spans of the agents
are disjoint and each agent makes connection sequentially one by
one.

\textit{One-mode projection of the fixed partition (place set)} is
a place to place (weighted) graph where two places are connected
by an edge if there is one common agent who visited/connected both
of the places. We denote this structure by ${\mathcal G}_P$. The
weight of an edge between two places in this projection denotes
the number of parallel edges between the two places via same or
different agents (see figure \ref{fig:BIPModel}(b)).

\textit{Thresholded one-mode projection}, is a simple graph
derived from ${\mathcal G}_P$ where the edges whose weights fall
below a certain threshold value $c$ in ${\mathcal G}_P$ are pruned
out. We denote this construction by ${\mathcal G}_{P}^{*}$ (see
figure \ref{fig:BIPModel}(c)).

\textit{Time varying threshold}: In general the value of the
threshold $c$, can be taken to be a constant. But, it is apparent
that as time grows (i.e., as more agents join the system), the
structure ${\mathcal G}_P$ constructed from the \abin{} gradually
becomes a complete (weighted) graph. Thus, for a constant
threshold, the corresponding ${\mathcal G}_P^*$ also becomes a
complete (un-weighted) graph for sufficient growth of the \abin.
However, it has been shown in \cite{BIP-JSac} that, instead of
taking constant value, if the threshold value is computed by
multiplying the number of agents joining the system (i.e., $t$) by
a constant (say \vthr,  may be fraction also), i.e. $v~\times~t$,
then certain properties of the resulting structure ${\mathcal
G}_P^*$ converge and become time invariant, for a given value of
$v$. We call this special kind of threshold value \vthr{} as
\textit{time varying threshold}.


\noindent{\bf Degree distribution in \GPstar{}:} The work
\cite{BIP-Saptarshi} shows that the preferential growth of an
\abin{} can be understood as a variant of the well known P\'{o}lya
urn model. Exploiting the exchangeability property of this
P\'{o}lya urn model and applying de Finetti's theorem, they show
that a single realization of the evolution of an \abin{} can be
mathematically understood in two steps - (a) first, a parameter,
sampled from a Dirichlet distribution is preassigned with each of
the place nodes in the fixed partition. We call this preassigned
parameters as the attractiveness parameters and denote it by
$\theta_i$ for node $i$. (b) Next, the agents are assumed to
select a place node $i$ through a Bernoulli process with success
rate equal to the attractiveness of $i$. The marginal distribution
of the attractiveness parameters is a beta distribution. For
special initial condition where all place nodes have equal degree
1, this beta distribution has the parameter set (1, $N$-1).

Furthermore, based on the above results, the authors in
\cite{BIP-JSac} derive that the growth rate of the weight of the
edges between two nodes $i$ and $j$ in the one-mode projection of
such \abin, is asymptotically equal to the product of $\theta_i$
and $\theta_j$ and $(\mu_2 - \mu)$, i.e.,
\begin{equation}
\label{eq:product}
\lim_{t\to\infty} \frac{{W}(i, j)}{t} =
(\mu_{2} - \mu) ~\theta_{i} \theta_{j}
\end{equation}
\noindent where $\theta_i$ and $\theta_j$ are the values of the
attractiveness parameters associated with nodes $i$ and $j$,
respectively; $W(i,j)$ is the weight of the edge between nodes $i$
and $j$ and $\mu$ and $\mu_{2}$ are the first and second moments
of the distribution of the number of edges created by the nodes
entering the growing set $V$.

The degree of a place node $i$ in \GPstar{} for a
threshold value $c$ (= $v\times t$, at time $t$) will be the number
of edges that are connected with $i$ in ${\mathcal G}_P$ and have weights
larger equal $c$. Thus, exploiting equation \ref{eq:product}, the
work \cite{BIP-JSac} also derives the cumulative degree
distribution ($F_v(k)$) of \GPstar{} for a given time-varying
threshold $v$ as follows -

\begin{equation}\label{eq:degree}
F_v(k)=\left(1-\frac{v}{(\mu^2-\mu)x}\right)^{(N-1)},
x=1-\left(\frac{k}{N-1}\right)^\frac{1}{N-1}
\end{equation}

\textbf{Special property of ${\mathcal G}_P^*$:} A thresholded projection generally
consists of a number of components (see figure \ref{fig:BIPModel}). However, in case of a
preferentially grown \abin, due to a special distribution of the weights
of the edges in ${\mathcal G}_P$, an interesting property is observed
in the component structure of ${\mathcal G}_P^*$. This property relates
the degree distribution of ${\mathcal G}_P^*$ and its largest component
size. It (the property) is based on the following lemma.

\textit{Lemma $C$}: \textit{The nodes which are in the largest
component in \GPstar{} have strictly higher degree in \G{} than
the nodes which are outside the largest component in \GPstar}.

\textit{Argument}: In the work \cite{BIP-Saptarshi}, it has been
derived that the expected degree of a node $i$ ($e(i)$) of set
\Pl{} in \GPstar{} is equal to
\begin{equation}\label{eq:degreeResult}
(N-1)\left(1-\frac{v}{(\mu_2-\mu)\theta_i}\right)^{(N-1)}
\end{equation}
where \vthr{} is the applied time varying threshold, \att{i} is
the probability that an agent will connect to node $i$ (in
\cite{BIP-Saptarshi}, this has been defined as the attractiveness
of node $i$) and $\mu$, $\mu_2$ are respectively, the first and
the second moments of the distribution of the number of edges
created by the agent nodes entering the set \A{}. Let us consider
two nodes $i$ and $j$, where node $i$ has higher degree in
\GPstar{} than node $j$, i.e., $e(i)>e(j)$. Using result
\ref{eq:degreeResult}, we can write that

\begin{eqnarray*}\label{eq:degreeInequality}
(N-1)\left(1-\frac{v}{(\mu_2-\mu)\theta_i}\right)^{(N-1)}>\nonumber\\
(N-1)\left(1-\frac{v}{(\mu_2-\mu)\theta_j}\right)^{(N-1)}
\end{eqnarray*}

Manipulating this inequality, it can be easily proved that
$\theta_i>\theta_j$. This implies, if a node $i~\epsilon~P$ has
smaller degree than some other node $j~\epsilon~P$ in \GPstar,
then the attractiveness of node $i$ is lower than that of node $j$
(in \G). This attractiveness of a node $i$ ($\theta_i$), is a
property of the BNW \G{} and is directly proportional to the
degree of $i$ in \G{} (i.e., depends on how many agent's
connection a node could attract to itself in the past). Now, since
we assume that at $t$=$T$, the nodes outside the largest component
in \GPstar{} are degenerate, it simply follows that the nodes
inside the component have higher degree in the BNW \G{} itself (by
virtue of having higher degree in \GPstar) than the nodes outside
the component. This proves the claim. $\diamond$

In the following we present the special property in the form of a
theorem.

\textit{Theorem : For any given threshold value \vthr, with high
probability, the thresholded one-mode projection of a
preferentially grown \abin, contains a single connected component
and all other nodes remain in the degenerate state.}

\textit{Implication of the theorem}: The property indicates a
specific structure of the thresholded one-mode projection of a
preferentially grown \abin. This property also relates the number of
components and the size of the largest component in \GPstar{} as
follows. For any threshold value \vthr, if  the
number of components in the thresholded one-mode projection,
is ($C_v$) and the size (i.e., number of nodes) of the largest
connected component is ($G_b$), then with high probability,  $C_v$ + $G_b$ = $N$ + 1. where $N$ is the total
number of nodes in the graph.

The proof follows.


\textit{Proof:} Let us consider the \abin{} \G{} with set of
agents \A{} and the set of places \Pl. Recall that the one-mode
projection on the place set is \GP{} and the thresholded one mode
projection is \GPstar.

\textit{Base condition}: For \nagent=0, i.e., when no agent has
joined the system, the thresholded one mode projection will
contain \nplace{} isolated nodes. Therefore, the theorem is
trivially valid.

\textit{Induction hypotheses}: We assume that for \nagent=$T$, the
thresholded one mode projection satisfies the theorem.

\textit{Induction step}: We now prove that the theorem still holds
true (with high probability) when agent number \nagent=$T$+1
completes all its connections.

As the theorem is satisfied for \nagent=$T$, we can say that after
$T$ agents have joined the system, in \GPstar, there is a single
connected component of size less or equal to \nplace{} and the
other nodes are isolated with degree zero.

Let us consider that the $(T+1)^{th}$ agent joins and creates
\navgcon{} number of connections. Let us focus on two subsets of
the nodes from set \Pl{} in \GPstar{} at this instant of time. One
subset (\SI) contains all the nodes with degree $u$ (in \G) and
they currently reside inside the largest component and the other
subset (\SII) contains the nodes outside the largest component
comprising all the nodes of degree $v$ (in \G) (from the Claim
$C$, it can be very easily proved that if some node of degree $d$
is outside the largest component, then all other nodes with the
same degree are also outside the largest component. Similarly, the
same is true for the degrees of the nodes inside the largest
component).

We denote the probability that a randomly selected node is of
degree $u$ and $v$ by \p{u} and \p{v} respectively. Therefore, the
number of nodes of degree $u$ and $v$, i.e., the cardinality of
the subset \SI{} and \SII{} are $Np_u$ and $Np_v$ respectively. As
already $T$ agents have joined the system, the probability that
the $(T+1)^{th}$ agent will select a $u$ degree node is
$\frac{u}{T}$ (due to fully preferential selection). Therefore,
out of $\mu$ connections of the $(T+1)^{th}$ agent,
$\frac{u\mu}{T}Np_u$ connections will be towards the subset \SI{}
and similarly $\frac{v\mu}{T}Np_v$ connections will be towards the
subset \SII.

Now, we consider three subsets of edges in the \GP{}: \EI, \EII{}
and \EIII. An element of set \EI{} connects two nodes both of
which are element of subset \SI{} (i.e., connects two nodes both
of which are inside the largest component). An element of set
\EII{} connects two nodes where one is in subset \SI{} and the
other is in subset \SII{} (i.e., connects elements one from within
and one outside the largest connected component). The elements of
set \EIII{} connects two elements both of which are elements of
the subset \SII{} (i.e., connects two elements outside the largest
component). The maximum number of elements possible in \EI, \EII{}
and \EIII{} are respectively $\binom{Np_u}{2}$,$Np_u\times Np_v$
and $\binom{Np_v}{2}$. These maximum number of possible edges can
be thought as holes where edges which are newly created due to the
joining of $(T+1)^{th}$ agent, get uniformly distributed. As a
result of creating $\frac{u\mu}{T}Np_u$ and $\frac{v\mu}{T}Np_v$
edges by the $(T+1)^{th}$ agent, with the elements of the subset
\SI{} and \SII{} respectively (in \G{}), the number of edges that
will be created in \GP{} are $\binom{\frac{u\mu Np_u}{T}}{2}$,
$\frac{u\mu Np_u}{T}\times \frac{v\mu Np_v}{T}$ and
$\binom{\frac{v\mu Np_v}{T}}{2}$ respectively (assuming a node of
\A{} creates at maximum one connection to a node of \Pl).
Therefore, the rate of increment of the weights of the edges of
the elements forming the subset \EI, \EII{} and \EIII{} are
respectively $\frac{\binom{\frac{u\mu
Np_u}{T}}{2}}{\binom{Np_u}{2}}$, $\frac{\frac{u\mu Np_u}{T}\times
\frac{v\mu Np_v}{T}}{Np_u\times Np_v}$ and
$\frac{\binom{\frac{v\mu Np_v}{T}}{2}}{\binom{Np_v}{2}}$. Let us
denote these quantities as \RI, \RII{} and \RIII{} respectively.
Simplifying these rates we get $r_1$ is $\mathcal{O}(u^2)$, $r_2$
is $\mathcal{O}(uv)$ and $r_3$ is $\mathcal{O}(v^2)$.

Using the claim $C$, we can say that $u>v$ because, the nodes of
\SI{} are inside the largest component and the nodes of set \SII{}
are outside the largest component. Comparing the rates with each
other, the following strictly descending order is obtained :
$r_1>r_2>r_3$.

From this order we can conclude that the edges in \GP{} can be
arranged in a strictly descending order (with high probability) in
the following way based on the rate of increment of their weights:
\textit{edges connecting two nodes both of which are inside the
largest component}, \textit{edges connecting two nodes exactly one
of which is in the largest component}, \textit{edges connecting
nodes both of which are isolated}. Therefore, with high
probability, those edges which connect two nodes - one outside and
another inside the largest component, will cross the threshold
weight sooner than the edges which connect two nodes both of which
are outside the largest component. As a result, the theorem still
holds true after the $(T+1)^{th}$ agent completes its connections
(with high probability). $\diamond$\\
\newline

\begin{figure*}[htbp]
\begin{center}
\includegraphics[angle=0,width=0.6\textwidth]{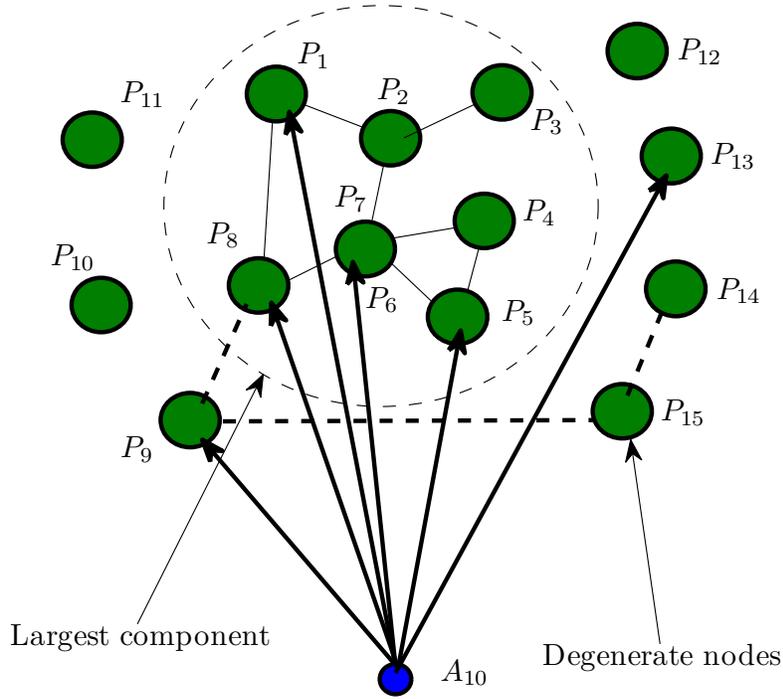}
\end{center}
\caption{Schematic diagram of a possible situation at the
$10^{th}$ time step in the evolution process in a BNW (\G) when
$10^{th}$ agent joins the system. Place $P_1$...$P_8$ have already
formed the single large component (encapsulated by dashed circle)
and the rest nodes $P_9$ to $P_{15}$ are in a degenerate condition
in the \GPstar. The solid lines show the connections after
applying threshold (i.e., in \GPstar) and the dashed lines show
the edges before thresholding in the one mode projection (i.e., in
\GP). At time step $T$=10, agent node $A_{10}$ arrives and creates
6 connections with the nodes in the place set (the lines with
arrow head shows these connections). Nodes $P_1$, $P_8$ and $P_6$
(inside the largest component) have degree 5 each and nodes $P_9$
and $P_{15}$ (outside the largest) have degree 2 each in \G. Now,
putting the values of $u$, $v$, $T$ and $\mu$ as 5,2,10 and 6
respectively, in the corresponding expressions of the rates
(derived in the proof), it can be calculated that (approximately)
$r_1$=9, $r_2$=3.6 and $r_3$=1.44. Hence, with high probability,
the edge connecting $P_8$ and $P_9$ in \GP{} (average edge weight
increment rate is 3.6 per agent) will acquire enough weight to
cross the threshold weight much earlier than the edge connecting
$P_9$ and $P_{15}$ (average edge weight increment rate is 1.44)
and therefore, formation of a separate component does not occur
and thus, with high probability the theorem still holds true after
agent node $A_{10}$ completes all its connections.}
\label{fig:ProofPicture}
\end{figure*}

Figure \ref{fig:ProofPicture} shows a possible situation in the
evolution process of the BNW when 9 agents have already joined the
system, a single large component has already been formed and the
$10^{th}$ agent joins.

\noindent{\bf Size of the largest component:} Exploiting this
property we calculate the size of the largest connected component
of ${\mathcal G}_P^*$ as the fraction of the nodes having degree
larger or equal to 1. We arrive at this quantity from the
cumulative degree distribution of \GPstar{} (equation
\ref{eq:degree}) as follows -
\begin{equation}
\label{eq:Gbformula} G_b=N\times\left[ 1-
\left(\frac{\sqrt[N-1]{(N-1)}}{\sqrt[N-1]{(N-1)}-1}\right)\times\left(\frac{v}{\mu^2-\mu}\right)\right]^{N-1}
\end{equation}


It can be inferred from \cite{BIP-JSac}, that the $F_v(k)$, for a
given value of \vthr{} also bears time-invariant characteristics.
\giantbnw{} is directly calculated from $F_v(k)$, hence is also
independent of $t$. It can be concluded that the \giantbnw{}
converges to a specific value for a given value of \vthr; in other
words it becomes independent of the number of agents that joined
the system.

\subsection{Correspondence between \abin{} and DTN}

In both the BNW as well as the DTN domain, the parameters \nplace,
\navgcon{} and \nagent{} have the same significance. In the DTN
setup, the flow of the message between two places is controlled by
the applied refresh probability \buftime. The higher the value of
\buftime, the lower is the probability that a message will
effectively get conveyed between a pair of places. We model this
refresh probability using the parameter \vthr{} of the BNW. Hence,
we estimate the coverage of information dissemination achieved
under a certain value of \buftime, i.e., \giantdtn{} by
calculating the size of the largest component of the ${\mathcal
G}_P^*$ in the corresponding BNW.

The parameters of the DTN and their relationship with those of the
BNW are summarized in table \ref{table:summary}.

\begin{table}[htbp]
\caption{Relationship between the parameters in DTN and the
corresponding BNW}
{\begin{tabular}{|p{1cm}|p{2cm}|p{2cm}|p{2cm}|}
\hline
Type        &   DTN           & BNW       &Remarks\\
\hline
Parameters  & Agents($t$)     & Agent partition ($t$)   & Growing\\
\hline
            & Places($N$)     & Place partition ($N$)   & Fixed and finite\\ \hline
            & Number of place an agent visits ($\mu$) & Number of connections an agent creates with different places ($\mu$) & Constant (can be taken from some specified distribution also)\\
            \hline
            & Buffer refresh probability ($p$) & Threshold varying with $t$, ($v$) & \\
            \hline
Observable  & Number of places where the message could reach under the dissemination process (\giantdtn) & Size of the largest component of the thresholded one-mode projection  (\giantbnw)& These quantities should match\\
            \hline
\end{tabular}}
\label{table:summary}
\end{table}

It is apparent that the two key parameters \buftime{} and \vthr{}
create a bridge between the two domains - DTN and BNW. Hence, we
also focus on understanding the functional relationship between
\vthr{} and \buftime.

\begin{figure*}[htbp]
\begin{center}
\includegraphics[width=\textwidth, height = 5cm]{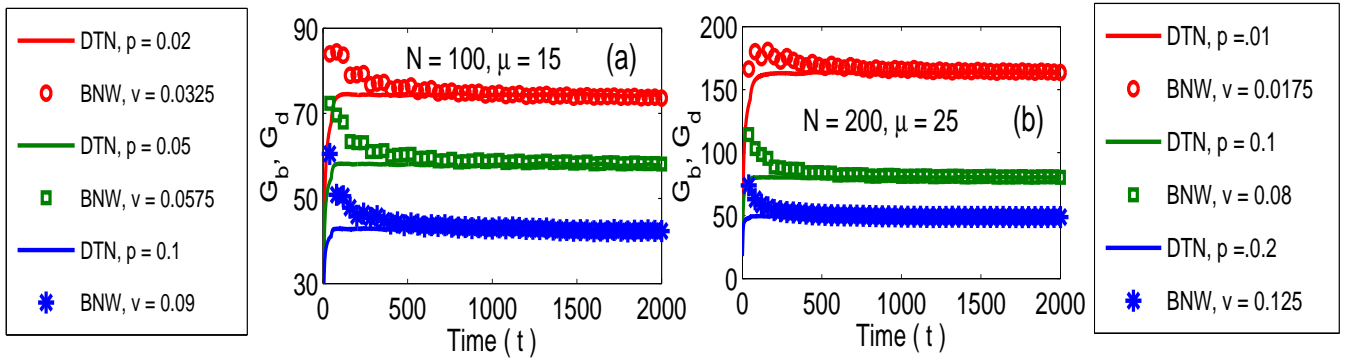}
\end{center}
\caption{Comparison of the size of the largest component
(\giantbnw) in the thresholded one-mode projection of the BNW and
the number of places in DTN where the message being disseminated
is found, i.e., coverage (\giantdtn). We consider the arrival of
2000 agents in both DTN and BNW.  The two parts (a) and (b) of the
figure show the results for two different combinations of
\nplace{} and \navgcon. The results shown are the average over 100
runs of simulation.} \label{fig:superimposed}
\end{figure*}

\textit{Empirical validation}: We simulate the time evolution of
both \giantdtn{} as well as \giantbnw{} for many different
parameter combinations. Figure \ref{fig:superimposed} shows few
such sample cases for different combinations of the other two
parameters \nplace{} and \navgcon. It can be seen that the time
series of these two quantities perfectly overlap for different
choices of $(v,p)$ pairs.

\begin{figure*}[htbp]
\begin{center}
\includegraphics[height=5cm, width=0.7\textwidth]{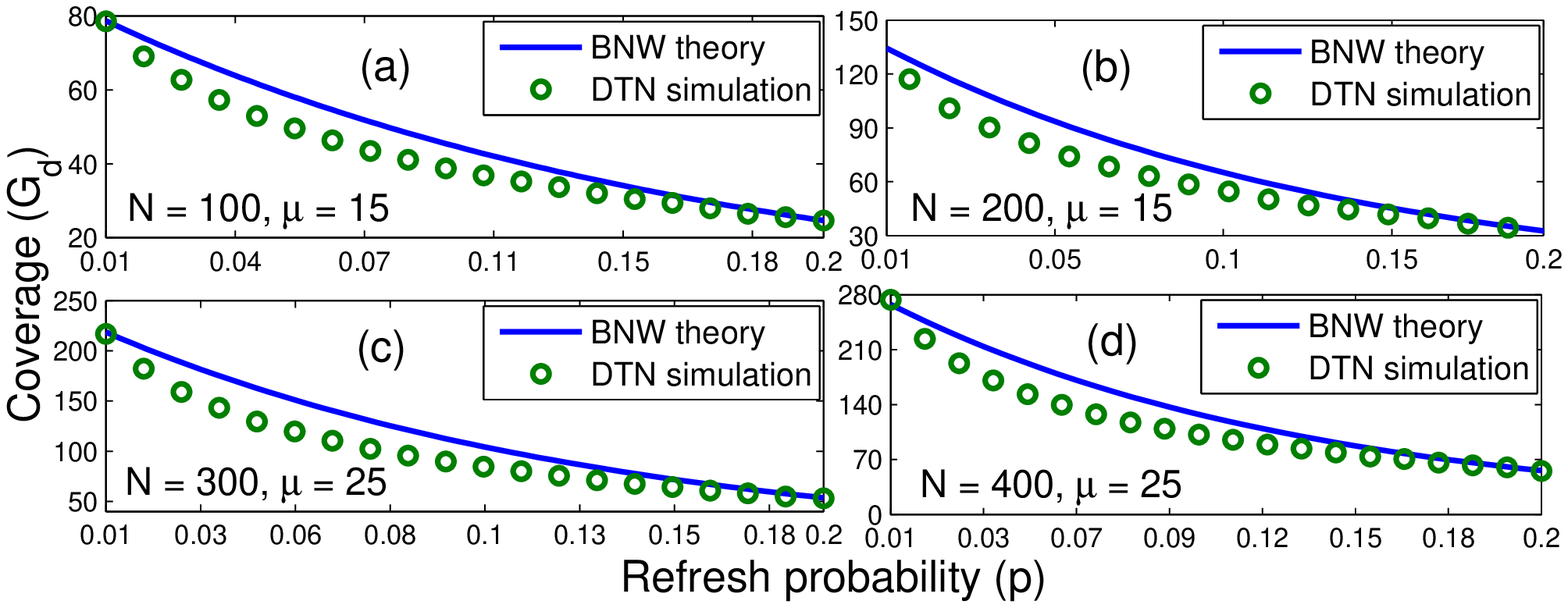}
\end{center}
\caption{Comparison of the simulation of DTN and the theoretical
predictions obtained from the corresponding BNW after proper
linear re-scaling of the range of the values of \vthr{} and
\buftime. Different parts of the figure shows the same result for
four different combinations of \nplace and \navgcon. The results
shown are the average over 100 runs of simulation.}
\label{fig:alpha-N-mu}
\end{figure*}

Through experimentation we observe that, a linear relationship of
the form $v=mp+c$, (where $m$ and $c$ are constants) exists in
between \vthr{} and \buftime. Figure \ref{fig:alpha-N-mu} shows
the comparison between the DTN simulation and the theoretical
predictions from BNW. The theoretical results of BNW are derived
in the following manner - at first, we derive the size of the
giant component ($G_b$)  for a whole range of $v$ value and then
select that portion of the result (range of $v$) where the highest
and the lowest value of $G_b$ perfectly overlap with the $G_d$
corresponding to the practical refresh probability ($p$) range
(0.01 to 0.2). This portion is then plotted with the DTN result by
linearly stretching the $v$ value to match the $p$ range. From the
graph it is evident that just a linear transformation of $v$ value
explains DTN coverage with high accuracy.

With this evidence, we now proceed to understand analytically the
influence of various parameters on the performance of DTN (i.e.,
coverage).

\section{Influence of the model parameters on coverage}\label{sec:insights}

In order to understand the role of the system parameters, we
exploit the linear relationship between $p$ and $v$ - we replace
\vthr{} by \buftime{} in equation \ref{eq:Gbformula}, we also
replace $G_b$ by $G_d$. Furthermore, using standard simplification
techniques, the equation \ref{eq:Gbformula} changes to the
following -

\begin{equation}
\label{eq:Gbformulasimple}
G_d=N-kN^{2}\frac{p(N-1)}{\mu(\mu-1)}
\end{equation}

\noindent where $k$ is a constant.

In the following we present several significant insights drawn
from the above mentioned simple formula.

\subsection{Effect of the basic \abin{} parameters}

\begin{figure*}[htbp]
\begin{center}
\includegraphics[width=0.75\textwidth, height = 9cm]{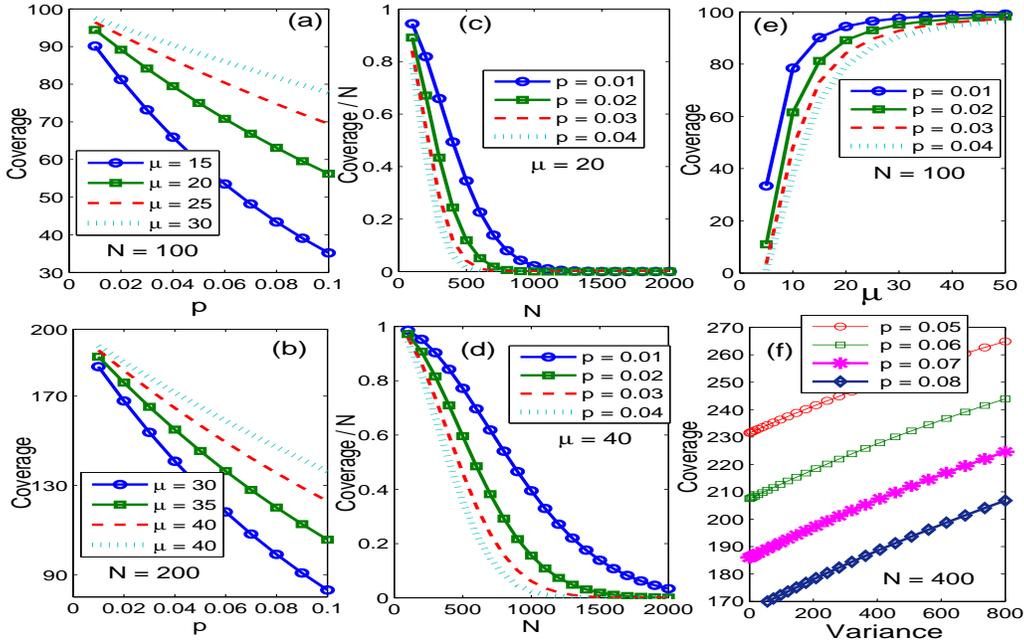}
\end{center}
\caption{Parts (a) and (b) show the plot of the relationship
between the refresh probability and the coverage in DTN, for
different combinations \nplace{} and \navgcon. Parts (c) and (d)
show dependence of the fraction of the nodes covered in DTN, i.e,
$\frac{coverage}{N}$ on the total number of places for different
combinations of \navgcon{} and \buftime. Part (e) shows the
relationship between the coverage in DTN and \navgcon{} for
\nplace=100 and different values of \buftime. All these presented
relationships are derived from equation \ref{eq:Gbformulasimple}.
Part (f) shows the relationship between the coverage and the
variance of the distribution of the number of places visited by
the agents for \nplace{} = 1000 and few different values of the
\buftime. We consider an uniform distribution of the number of
places visited by the agents. We set the value of average of this
distribution (=\navgcon) at 50 and vary the range of the value
(such as [1,99], [11, 89], [21,79] etc.) to vary the variance. For
each different variance we calculate the achieved coverage from
equation \ref{eq:Gbformulasimple}.} \label{fig:insights}
\end{figure*}

\indent\textbf{Effect of \buftime} : For a certain fixed number of
common places (\nplace) and for a fixed number of visits
(\navgcon) by the agents, the achieved coverage in DTN
decreases linearly with the refresh probability. Figures
\ref{fig:insights}(a) and (b) show few such sample results
bringing out this linear relationship between the coverage and the
\buftime{} (in the admissible range) for different choices of
\nplace{} and \navgcon.

\indent\textbf{Effect of \nplace}: For a fixed number of visits by
the agents and for a given refresh probability, the possibility of
full coverage in DTN decreases at a rate proportional to the cubic
power of the number of common places (i.e., $N^{3}$), that is,
with the increase in size of $N$, it becomes very difficult to
have full coverage. To illustrate this in figures
\ref{fig:insights}(c) and (d), we plot the normalized coverage,
that is, Coverage/$N$ in the y-axis. As seen in the graph, the
coverage quickly diminishes with the increase in \nplace.

\indent\textbf{Effect of \navgcon}: For a certain fixed number of
common places and for a given refresh probability, the achieved
coverage in DTN increases at a rate proportional to $\mu^{2}$.
Figure \ref{fig:insights}(e) presents few such sample results
showing this relationship between \giantdtn{} and \navgcon. It can
be seen from this figure that, up to a certain value of \navgcon,
the achieved coverage \giantdtn{} increases rapidly (provided all
other parameters remain fixed). However, beyond that, coverage
hardly increases.

Furthermore, it can be shown that for a generalized distribution
of the number of visits of the agents in the system (let us denote
it by $F(x)$), the denominator ($\mu^2$ - $\mu$) in equation
\ref{eq:Gbformulasimple}, can be replaced by
($\sigma^2+\psi(\psi-1))$ where $\psi$ and $\sigma$ are the
average and the standard deviation of $F(x)$ (as shown in
\cite{BIP-Saptarshi,BIP-JSac}). Hence the inferences are (a) the
achieved coverage only depends on the first and the second moment
of $F(x)$ and (b) for a fixed value of $\psi$, if the number of
agent visits are more diverse, (i.e., $F(x)$ has higher variance),
then the achieved coverage in the dissemination process will be
higher, i.e., the larger $F(x)$ varies, the higher is the
coverage. Figure \ref{fig:insights}(f) presents few such results
showing the relationship between the variance of $F(x)$ and the
achieved coverage (detailed in the caption of the figure).

\subsection{Effect of mobility related parameters}
We also consider two other parameters related to the mobility of
the agents - (a) randomness in selecting new places ($\delta$) (b)
propensity of the agents to visit nearer places, termed as
clustering exponent ($\alpha$). These two terms have been
discussed in section \ref{subsec:placeselection} and equation
\ref{eq:pref_rand_probability}.

\begin{figure*}
\begin{center}
\includegraphics[height = 7cm, width =0.8\textwidth]{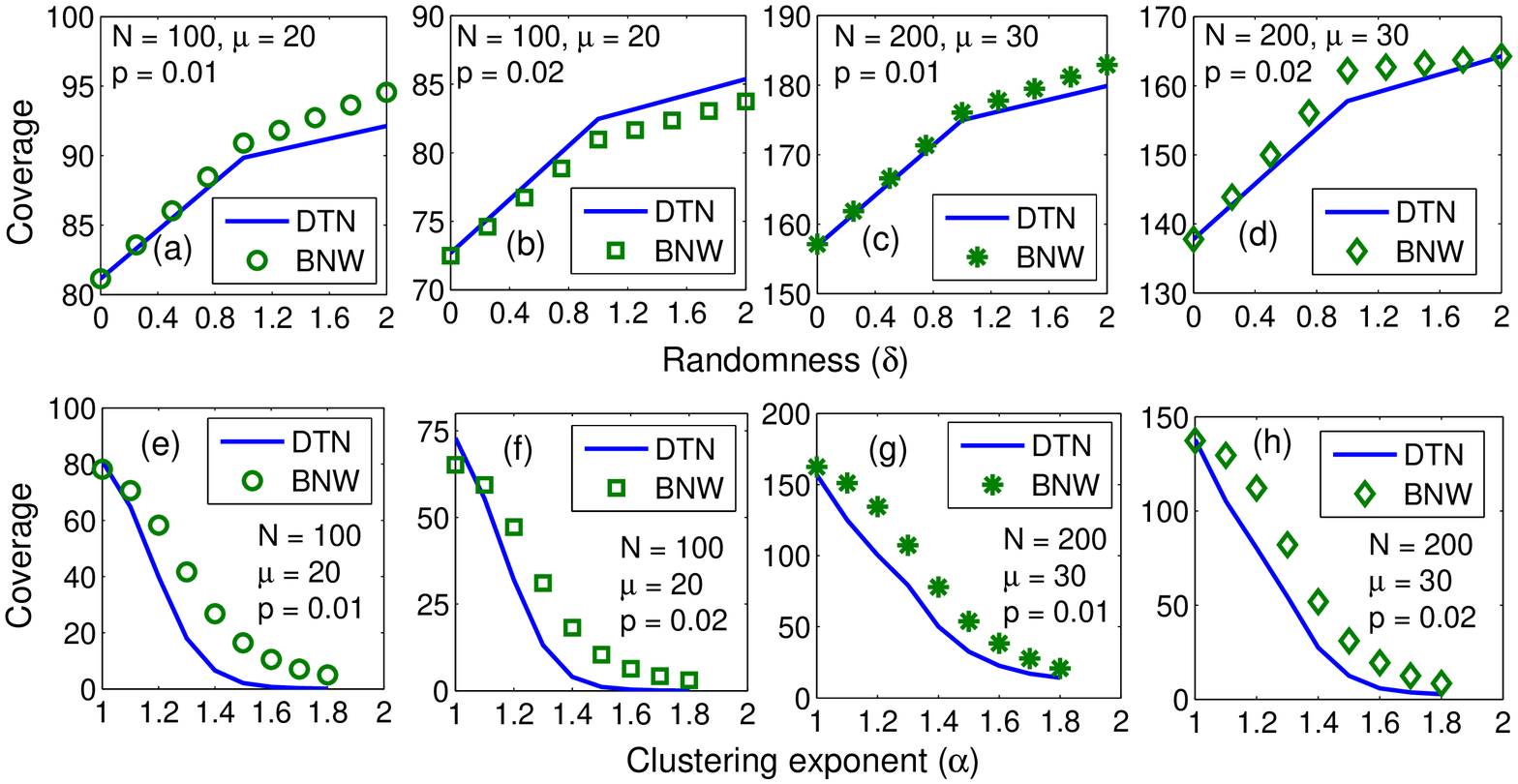}
\end{center}
\caption{Parts (a) - (d) show the plot of the achieved coverage in
the simulation of message dissemination in DTN, for several
different values of \rand, under two different refresh
probabilities and two different combinations of \nplace{} and
\navgcon. Parts (e) - (h) show the plot of the achieved coverage
in DTN, for several different values of \clust, under two
different refresh probabilities and two different combinations of
\nplace{} and \navgcon. The simulation results obtained from the
BNW have been also shown. All the simulation results were computed
by averaging over 1000 runs.} \label{fig:clustrand}
\end{figure*}

\indent\textbf{Effect of randomness, \rand} : We find that the
achieved coverage increases noticeably for the initial values of
\rand{} (between 0 and 1) and changes very slowly for higher
values. Figures \ref{fig:clustrand}(a) to (d), show some sample
results for few different combinations of \nplace, \navgcon{} and
\buftime. The reason for the improvement in coverage with the
slight increase in the randomness is that with randomness, people
tend to visit many other places with higher frequency in
comparison to the pure preferential case.

\indent\textbf{Effect of clustering exponent, \clust} : We observe
completely opposite effect of \clust{} on the achieved coverage.
We find a drastic decrease in \giantdtn{} with even a very small
increase in the \clust. Figures \ref{fig:clustrand}(e) to (h)
show some sample results for few different combinations of
\nplace, \navgcon{} and \buftime. This happens due to the fact
that, for higher value of \clust, people tend to visit in a more
clustered fashion, i.e., few places start getting visited even at
higher frequencies while the other places get visited with lesser
frequencies in comparison to the pure preferential case.
Subsequently a super-preferential attachment dynamics emerges;
hence the number of places preserving the message
 drastically decreases.

\section{Related Work}\label{sec:RelatedWork}

Bipartite networks and their projections have been employed as a
model in various diverse fields, e.g., personal recommendation
\cite{2014_arxiv_nie_bnw_recommendation,2007_pre_zhou_bnw_personal_recommendation,2013_plosone_zeng_bnw_recommendation},
analysis of social networks
\cite{2013_epjb_liu_bnw_socialinfluence,2013_socialnetworks_opsahl_bnw,2013_socialcomputing_yo_bnw_dating},
protein structures \cite{2002_science_maslov_protein}, linguistics
\cite{BIP-choudhury-2010-81} etc. In the current work we use an
evolving bipartite network \cite{BIP-Saptarshi} to model the
message dissemination process in buffer augmented DTN. On the
other hand, the spreading processes also have been studied
immensely using various models such as SI, SIS, SIR, SIRS
\cite{1991_book_oxfordpress_anderson_may_disease_spreading_math,2008_princeton_keeling_infectious_disease_human_animals}
etc. Even recently bipartite networks have been studied under such
frameworks
\cite{2013_pre_hernandez_bnw_epidemic_thresholds,2013_scientificreport_huang_bnw_cascading}.
However, message dissemination in DTN being a computer-science
application specific problem, the place coverage as studied in
this paper, is more significant in this context than various other
critical issues which are generally analyzed in epidemiology. In
the following we briefly describe the existing works in this
specific direction.

In the communication network literature, from the perspective of
information dissemination, the coverage, i.e., number of
distinctly visited nodes in any network, has been a significant
metric. Many works have been done on developing either efficient
algorithms or analyzing the maximum achievable
coverage~\cite{Probabilistic-Flooding-1,KRandomWalkHLarralde,CoverageMaximizationSubrata,probabilistic-flooding-2,probabilistic-flooding-3}
in both wired and wireless networks through appropriate modeling
of the undergoing process. However, similar work for DTN
(augmented with message buffers
\cite{DTN-info-disse-Throwbox,relay-point-active-2}) is largely
missing in the literature.

Most dissemination schemes proposed for DTN follow a
store-carry-forward paradigm in which the mobile devices carried
by the agents temporarily buffer the data and forward it to other
(appropriate) agents. Various strategies like epidemic routing
\cite{epidemic-base}, spray and wait protocols
\cite{spray-wait-density}, contact history-based protocols
\cite{contact-3} have been employed to enhance the speed and
amount of coverage. In addition, the efficiency of any such
strategy heavily depends on exploiting the patterns emerging from
the underlying mobility models. Various such models have been
proposed - from the simple random walk
\cite{HM-RANDOM-WAYPOINT-MOBILITY} to the more sophisticated
models such as \textit{self similar least action walk}
~\cite{HM-SLAW-DBLP-conf-infocom-LeeHKRC09}.

A majority of the past studies assume a suitable mathematical
abstraction of the real world DTN environment and observe the
important characteristics through simulations and analytics. This
is a very standard and well accepted approach adopted in the
literature \cite{epidemic-base,spray-wait-density} which has been
also the central policy of the current work. Several stochastic
techniques like epidemic spreading \cite{modeling-epidemic-1},
ordinary differential equations \cite{modeling-ode-2}, partial
differential equations \cite{modeling-pde-1} as well as Markov
models \cite{DTN-info-disse-ModellingDTN} have been applied to
gain the necessary analytical insight. However, the use of
store-carry-forward paradigm in the dissemination processes
adopted for DTN in conjunction with the complex human mobility
models, make it extremely difficult to produce {\em neat} results
using such traditional mathematical tools. All the above mentioned
works in the direction of analyzing coverage, either use very
simplistic setup or ultimately land up in very complicated {\em
open} equations where the relationships among the key parameters
of the system cannot be easily understood from these equations.
The work presented in this paper is significant in the sense that
we come up with simple closed form equations through clever
modelling of the complex process.

Recently, Gu et al \cite{SimilarWork-recent-journal} have
addressed this issue in similar lines as those presented in this
paper. They have perceived the use of message buffers as an
instance of bio-inspired methods (e.g., pheromone or footprint).
Using discrete Markov chain based modeling, they analyzed the
importance of buffer time as well as the preferences of visiting
different places by the mobile agents. They studied the impact of
these two crucial system features on the latency and the message
delivery ratio of the dissemination process in the network. In our
work, we specifically emphasize on the inherent bipartite nature
of the ``throwbox augmented DTN'' and thus bring forward the fact
that instead of starting from scratch, the existing theories of
the bipartite network can be used (with necessary modifications)
to analyze the coverage problems related to DTN. In
\cite{DTN-buffer-mobiopp} we introduced this methodology for the
first time and reported certain initial results. The settings
assumed there have been largely artificial where neither real-life
traces, clustered mobility models etc. have been considered. We
also assumed theoretical setups like buffer time, i.e., the number
of time steps for which a given message is stored in a message
buffer (denoted by $b$), to be a globally fixed quantity. However,
here we adopt the buffer refresh probability which happens to be
more easily implementable (hence practical), maintainable and cost
effective solution. Moreover, we also demonstrate here our
detailed study on how the coverage in DTN gets affected depending
on various issues related to the mobility of the agents -like
their life spans, activities, variation in the place selection
strategy.

\section{Conclusion}
In this work we introduce a new mathematical tool (bipartite
network) to analyze {\em place coverage} in the process of
information dissemination through indirect message transfer in
DTN. We show that the size of the largest component of the
thresholded projection of an appropriately constructed bipartite
network, can be used to accurately measure {\em place coverage}.
We also demonstrate a straightforward way of calculating the size
of the largest component from the degree distribution. Moreover,
from the mathematical equations we derive several simple insights
regarding the complex process of message spreading in throwbox
augmented DTN. One of the most significant non-intuitive results
is that {\em place coverage} converges and does not increase
beyond a certain level for a given refresh probability ($p$) and
this level is significantly lower than full coverage for even a
very small value of $p$.  The lesson learned is that if the number
of places increases or the variation of agent activity is not very
high or if the agents have the tendency to roam within only
popular places, even the introduction of throwbox as a backbone
infrastructure may not be sufficient. 

\section*{Acknowledgment}
This work was funded by project BDT under \textit{Department of
Information and Technology} (DIT), Govt. of India. S. S. thanks
Tata Consultancy Services (TCS) Pvt Ltd and Samsung India Pvt Ltd,
for financial assistance.

\ifCLASSOPTIONcaptionsoff
  \newpage
\fi



%

\bibliographystyle{IEEEtrans}
\bibliography{DTN-BNW-TNSE}


%








\end{document}